\documentclass[useAMS,usenatbib]{mn2e}
\usepackage{epsfig}
\usepackage{graphicx}

\begin{document}
\voffset=-0.5 in

\title[A tale of two populations: RRATs and XDINSs] {A tale of two
populations: Rotating Radio Transients and X-ray Dim Isolated Neutron
Stars}

\author[S.B. Popov, R. Turolla, A. Possenti]{S.B. Popov$^{1}$,
R. Turolla$^2$ and A. Possenti$^3$
\thanks{E-mail: polar@sai.msu.ru
(SBP); turolla@pd.infn.it (RT); possenti@ca.astro.it (AP)}\\
$^1${Sternberg Astronomical Institute, Universitetski pr. 13, Moscow
119992, Russia} \\
$^2$ {Department of Physics, University of Padova,
via Marzolo 8, I-35131, Padova, Italy}\\
$^3$ {Istituto Nazionale di
Astrofisica, Osservatorio Astronomico di Cagliari, loc. Poggio dei
Pini, Strada 54, I-09012 Capoterra, Italy} }
\date{Accepted ......  Received ......; in original form ......
%(MNRAS, 2005, in press)
      }
%\pubyear{2005}

%\shorttitle{Popov, Turolla \& Possenti}{A tale of two populations:
%RRATs and XDINSs}

\maketitle

\begin{abstract}
We highlight similarities between recently discovered Rotating Radio
Transients and X-ray Dim Isolated Neutron Stars.  In particular, it is
shown that X-ray Dim Isolated Neutron Stars have a birthrate
comparable to that of Rotating Radio Transients. On the contrary,
magnetars have too low a formation rate to account for the bulk of the
radio transient population. The consequences of the recent detection
of a thermal X-ray source associated with one of the Rotating Radio
Transients on the proposed scenarios for these sources are also
discussed.
\end{abstract}


%\newpage

\section{Introduction}
\label{intro}

The remarkable discovery of a new population of transient
radio-sources in the Parkes Multibeam Survey (dubbed Rotating RAdio
Transients, RRATs) has been announced very recently \citep{mcl2005}.
A total of 11 sources have been detected so far and they are
characterized by the emission of bursts with duration $\sim 2$--30
ms. The maximum burst flux density is $\sim 0.1$--4~Jy and the mean
interval between bursts is $\sim 4$~min--3~hrs.  Although no
periodicities were found using standard FFT searches, time difference
analysis revealed periods in the range 0.4--7~s in 10 out of 11
sources with no signature for binarity. In three cases the period
derivative $\dot P$ was measured: the inferred magnetic fields and
ages are in the $10^{12}$--$10^{14}$~G and 0.1--3 Myrs range,
respectively. The sources appear concentrated towards the Galactic
plane and the inner Galaxy, like the normal young pulsar
population. Although their nature is still mysterious, a natural
inference is that we are observing a subset of the population of the
Galactic isolated neutron stars, with radio properties much at
variance with those of ordinary radio-pulsars (PSRs).

In principle, RRATs might be a completely new, disparate class of
isolated neutron stars, or they may be linked to an already classified
population. In fact, on the basis of the similarities in the measured
periods and period derivatives, \cite{mcl2005} pointed out a possible
association between RRATs and the anomalous X-ray pulsars/soft
$\gamma$-repeaters (AXPs/SGRs, the so called magnetar candidates;
e.g. \citealt{wt2004}), or between RRATs and the X-ray dim isolated
neutron stars (XDINSs). The latter are a small group of seven
radio-quiet, close-by X-ray pulsars, characterized by a thermal
spectrum with typical temperatures $kT\la 100$~eV and periods $P$ in
the range $\sim 3$--11~s (e.g. \citealt{hab04}). The recent measure of
the spin period derivative $\dot P$ in two sources gives magnetic
fields of $\sim (2-3)\times 10^{13}$~G and ages $\sim 1-2$~Myr
(\citealt{kvk2005}, \citealt{kvk2005a}).

The possibility that RRATs are far away XDINSs appears worth of
further consideration in view of the many, striking similarities
between the two populations: values of $P$ and $\dot P$, estimated
ages, spatial distribution, lack of pulsed radio emission. The
very recent discovery of a faint X-ray source (CXOU
J181934.1$-$145804) with soft thermal spectrum ($kT\sim 120$~eV;
\citealt{rey2005}) in positional coincidence with the RRAT
J1819-1458 lends further support to this option.

XDINSs have been searched for pulsed radio emission with negative
results (\citealt{bj1999}; \citealt{johns2003}), but only in one case
(RX J0720.4$-$3125; \citealt{bradley06}) the signal
analysis was sensitive to radio bursting activity. \cite{m05} reported
the detection of pulsed (probably not bursting) radio emission at very
low frequency from RX J1308.6+2127. If confirmed, the detection may be
related to the fact that this source is probably the only orthogonal
rotator among XDINSs. This geometry is inferred from the double-peaked
lightcurve in the X-ray band, and, interestingly enough, similar
properties are shared by the AXP 1E 2259+586, also possibly seen at
very low radio frequency \citep{m05}. A systematic multi-frequency
search for radio bursts from XDINSs is presently under way (Possenti
et al. in preparation) and, if positive, it would strongly suggest an
association between the two types of NSs.

In this Letter we first explore (\S~\ref{birthrates}) the possible
links between RRATs and SGRs/AXPs or XDINSs by a comparative study of
the properties of the three classes of isolated neutron stars,
focusing in particular on their birthrates. Consequences of the
discovery of an X-ray counterpart to RRAT J1819-1458 \citep{rey2005}
on the nature of RRATs are discussed in \S~\ref{discus}.

\section{Birthrates of different types of neutron stars}
\label{birthrates}

\subsection{PSRs and RRATs birthrate}
\label{rrat-psr-rates}

\cite{mcl2005} estimated the number of RRATs to be $N_{\rm RRAT}\sim
4\times 10^5$ while current estimates place the number of Galactic
PSRs in the range $7\times 10^4\la N_{\rm PSR}\la 1.2\times 10^5$
\citep{vr2004}. Although the determination of $N_{\rm RRAT}$ is still
preliminary, this would imply that RRATs outnumber active radio-pulsars
by a factor $\sim 3-5$. However, drawing the conclusion that being a
RRAT is a more common nature for a neutron stars than being a PSR
would be incorrect. The duration of the two phenomena, in fact, may be
substantially different and a better measure of the relative abundance
of RRATs and PSRs is obtained comparing the birthrates of the two
classes of neutron stars.

Recent investigations give for the PSR birthrate a value
$\beta_{\rm PSR}\sim 0.01$--$0.02\, {\rm yr}^{-1}$
(\citealt{vr2004}; see also \citealt{l2005} and \citealt{fk06}).
Likewise $N_{\rm PSR}$, the PSR birthrate depends on the adopted
models for pulsar beaming and Galactic electron distribution and
the assumed estimates of both quantities tend to produce lower
limits on the true population (see again \citealt{vr2004}). The
average active pulsar lifetime (i.e. the characteristic time
required to reach the death line) may be estimated as $\tau_{\rm
PSR}\approx N_{\rm PSR}/\beta_{\rm PSR} \sim 5\times 10^6\, {\rm
yr}$. Despite the uncertainties, $\tau_{\rm PSR}$ turns out to be
close to the typical values one obtains from the spin-down
formula, $\dot P\propto P^{2-n}$. This holds in the case of death
line for curvature radiation (star-centred dipolar field) for both
vacuum gap and space-charge-limited flow \citep{zh2000}, and
adopting a braking index close to those observed in young PSRs
($1.4\la n_{\rm PSR}\la 2.9$; \citealt{lkg06}).

Due to the very limited amount of information presently available, an
estimate of RRATs characteristic lifetime $\tau_{\rm RRAT}$ is somehow
haphazard. The physical mechanism responsible for the emission of
radio bursts is still unclear and, so far, there is no proof that it
is similar to that of canonical pulsar radio emission (see
\citealt{z06} for a first discussion on the RRAT nature). As a
consequence, the region in the $P$--$\dot P$ diagram in which the
conditions for burst emission are fulfilled may not coincide with that
in which pulsed emission is allowed, so the death lines for the two
types of radio activity may well be different. Moreover, the physical
mechanism responsible for the bursts may also influence the braking
index $n_{\rm RRAT}$ of RRATs. If $n_{\rm RRAT}$ differs substantially
from that of the bulk of the PSR population, the spin-down history of
RRATs may bear little resemblance to that of ordinary radio-pulsars.

This stipulated, it is interesting to note that the three RRATs
for which both $P$ and $\dot P$ have been measured lie above (or
close to) the radio-pulsar death line. Although the scanty
statistics may play a role here, this may suggest that minimal
conditions for triggering radio burst are shared with those for
normal radio-pulsar behaviour. But, again, none can ensure that
the typical time for reaching the death line is the same for PSRs
and RRATs. However, if RRATs are rotating neutron stars endowed
with pulsar-like magnetic fields, their magneto-dipolar spin-down
rate is expected to be similar to that of PSRs. Unless RRAT and
PSR distributions of initial periods and magnetic fields are
drastically different, one can then tentatively conclude that the
lifetime of the two populations is not much diverse. Even a
different braking index, e.g. produced by the bursting activity,
should not alter the main conclusion, since for $1\la n\la 4$ the
typical time required to reach the death line is within a factor
of a few. Given that $N_{\rm RRAT}=\gamma N_{\rm PSR}$, our
estimate $\tau_{\rm RRAT}\approx\tau_{\rm PSR}$ implies that
$\beta_{\rm RRAT}=\gamma \beta_{\rm PSR}$ with $\gamma > 3.$

\subsection{Magnetars birthrate}
\label{mag-rates}

\cite{ketal1998} estimated that magnetars are born with a rate
$\beta_{\rm SGR}\sim 0.001\, {\rm yr}^{-1}$.  This follows from the
estimated age of the soft repeater SGR 1806-20 ($\sim 8000$--10000~yr)
and from the upper limit on the number of SGRs in the Galaxy ($N_{\rm
SGR}\la 7$; see again \citealt{ketal1998}). AXPs are suspected to host
a magnetar too, and are about twice as numerous as SGRs. However,
since they are believed to be typically one order of magnitude older
than SGRs (see e.g. \citealt{wt2004}), the inferred birthrate for the
two types of sources is of the same order, $\beta_{\rm SGR}\sim
\beta_{\rm AXP}$.

Taken face value, $\beta_{\rm SGR}$ is about a few per cent of the
birthrate of radio-pulsars and obviously too low to be comparable
with the RRAT birthrate as given in \S~\ref{rrat-psr-rates}. We note
that the estimate by \cite{ketal1998} strongly depends on the
efficiency of SGR detection \citep{wt2004}. However, in order to
obtain a SGR birthrate equal to the rate of radio-pulsar formation,
one should assume a detection efficiency as low as 10\%, at variance
with the fact that each of the four confirmed SGRs produced a strong
flare during the period of active monitoring of these
sources\footnote{Three of these bursts are {\em bona fide} giant
flares.  The fourth, that from SGR 1627-41 observed on 1998 June 18,
was also very energetic, but did not show evidence for a pulsating
tail. Still, some authors classify this event as a giant flare
\citep{metal2004}.}. We also note that the stringent upper limits on
the detection of giant flares from extragalactic SGRs in the BATSE
archive \citep{lazzati2005, ps2006} provides further evidence in
favour of a small number of magnetars in the Galaxy.

Even regarding $\beta_{\rm SGR}\sim 0.001\, {\rm yr}^{-1}$ as an
absolute lower limit (not all magnetars in their youth may exhibit a
SGR-like activity), the birthrate of RRATs can hardly be reconciled
with that of magnetars. We conclude that some small fraction of RRATs
can perhaps be associated with magnetars, but not the bulk of the
population.

\subsection{XDINS birthrate}
\label{xdin-rates}

The population of XDINSs is small (up to now $N_{\rm XDINS}=7$, whence
the name {\it Magnificent Seven}), and comprises
only relatively young and close-by objects. This is mainly a
consequence of an observational bias, since intermediate-age,
radio-silent cooling neutron stars are very elusive sources.

Although it is difficult to get an exact estimate of $\beta_{\rm
XDINS}$, its order of magnitude can be easily derived. Only the
distance of the brightest source, RX J1856.5-3754 ($\sim 120-170\,
{\rm pc}$, \citealt{wlat2002}; D. Kaplan, unpublished) is well
determined. However, we can conservatively adopt a distance of
$D_{\rm XDINS}\la 400$ pc on the basis of the X-ray determination of
the column densities and on the upper limit for the size of the Gould Belt
(see below). Ages also are not well constrained, but, basing on the surface
temperatures, XDINSs cannot be older than $\tau_{\rm XDINS}\approx
1$~Myr, if they are standard cooling isolated neutron star (see
e.g. \citealt{yak99} and \citealt{page}). This is supported by the
recent measure of $\dot P$ in two sources, implying an age of $\sim
1-2$ Myr (\citealt{kvk2005}; \citealt{kvk2005a}). A first estimate of
the XDINS birthrate in the proximity of the Sun is then of order
ten per Myr, well consistent with the inferred supernova rate
in the Gould Belt, $\sim 20-30\,{\rm Myr}^{-1}$ \citep{g2000}.

In fact, as discussed by \cite{p03}, the local density of XDINSs is
expected to be higher with respect to the Galactic average since the
solar neighborhood is part of the Gould Belt, an association of young
stellar systems. Because of the comparatively large fraction of
massive stars in the Belt, the local neutron star formation rate is
higher.  The origin of XDINSs in young nearby stellar systems is also
supported by dynamical considerations based on their proper
motions (e.g. \citealt{kaplan2002}). In summary, the Gould Belt accounts
for about 2/3 of the observed XDINSs, so that the average Galactic
birthrate $\beta_{\rm XDINS}$ should be $\lambda_1\sim 3$ times lower
than the local one.  On the other hand, neutron star cooling
curves are mass-dependent (more massive stars cool faster, see again
\citealt{yak99}; \citealt{page}) and close-by NSs which are
slightly more massive than the observed seven may be too dim
by now to be detectable in X-rays.  In particular, we can expect up to
a dozen additional, dimmer isolated neutron stars of the same age and
within the same distance of the {\it Magnificent Seven} (see, for
example, \citealt{ad04, p04}).  These sources may be among the
unidentified ROSAT objects (\citealt{s99}; \citealt{rut2003}), and we
can expect them to be identified by on-going searches (see, for
example, \citealt{c2005, ag2005} for recent reports on the search for
new XDINS candidates). Hence the total number of XDINSs in the solar
neighborhood should be $\lambda_2N_{\rm XDINS},$ where $\lambda_2\sim
2-3.$

Extending the previous argument to the whole Galaxy gives $\beta_{\rm
XDINS}=[(\lambda_2N_{\rm XDINS}/\lambda_1)\times (R_{\rm disc}/D_{\rm
XDINS})^2]/\tau_{\rm XDINS} \approx 0.01\,{\rm yr}^{-1}$. In deriving
the previous expression we used the fact that XDINSs are a disc population
and took $R_{\rm disc}=15 \, {\rm kpc}$, $\lambda_2/\lambda_1\sim 1$. It
should also be mentioned that the local density of neutron stars
(disregarding the enhancement produced by the Belt) is lower than the
Galactic average, because of the exponential density decay (with typical
length-scale $\sim 3$ kpc; e.g. \citealt{k1991, fm1994})
in the Galactic disc.  If this is taken
into account, then $\beta_{\rm XDINS}$ may reach a few events per century
and it may well be that cooling, radio-silent isolated neutron
stars have a birthrate comparable to, or even higher than, that of
radio-pulsars. This is indeed the case for the local population of
XDINSs, as suspected already by \cite{nt99} and confirmed by
\cite{p00b} and by \cite{p03}. As these investigations have shown,
the average density of radio-pulsars is not high enough to explain the
number of observed XDINSs.

\subsection{Total birthrate of isolated neutron stars}
\label{ins-rates}

As a consistency check for our birthrate estimates of
\S\ref{rrat-psr-rates}, \S\ref{mag-rates}, and \S\ref{xdin-rates}, we
can compare the total birthrate $\beta_{\rm INS}$ of isolated neutron
stars (INSs) in the Galaxy with the observed rate of supernovae. In
fact, $\beta_{\rm INS}\ge\beta_{\rm PSR}+\beta_{\rm XDINS}+\beta_{\rm
SGR}.$ The formation rate of the magnetars can be safely neglected in
this calculation, since it is about a factor ten lower than both
$\beta_{\rm PSR}$ and $\beta_{\rm XDINS}.$ Hence, conservatively
assuming $\beta_{\rm PSR}\sim \beta_{\rm XDINS},$ and accounting for
the stated uncertainties, it turns out $\beta_{\rm INS} \ga
0.02-0.04$ events per yr. This figure is in comfortable correspondence
with both the recent determination of the rate of core collapse
supernovae in the Galaxy ($1.9\pm 1.1$ events per century:
\citealt{d06}) and with the estimate of the rate of supernovae which
can produce neutron stars ($\la 3$ events per century:
\citealt{fk06}).

\section{Discussion}
\label{discus}

In the previous section we have pointed out that among the known
Galactic INS populations, and with the possible exception of PSRs (see
below), XDINSs alone have a formation rate of the right order to
account for the large inferred number of RRATs. Although the idea that
RRATs are far away XDINSs appear promising in the light of presently
available information, other interpretations may be conceivable.

The Galaxy is the graveyard of as many as $10^8$ INSs which are
too old to be active as radio-pulsars. Despite old NSs may be
rejuvenated by the accretion of interstellar material (e.g.
\citealt{trev00} and references therein), such a population has
never been detected, most probably because accretion is impeded by
the star magnetic field, rotation and large spatial velocity. As
discussed by \cite{p00a}, a fraction of $\sim 0.1$--0.2\% of all
INSs is expected to be presently in the so-called supersonic
propeller stage.  This amounts to $\approx 10^5$ objects, a number
consistent with the estimate for RRATs. A supersonic propeller
might match the values of $P$ \citep{b05a,b05b} and $\dot P$
(\citealt{a2001} and references therein) observed in RRATs. The
emission of radio-bursts may be produced by particles accelerated
by magnetic reconnection in the elongated magnetotail (e.g.
\citealt{torop01}), although no detailed model has been worked out
yet.  Alternatively, RRATs may be just ordinary radio-pulsars,
with a particular viewing geometry, or pulsars from the ``death
valley'', as suggested by \cite{z06}.

The discovery of the X-ray counterpart for the RRAT J1819-1458 in {\em
Chandra} archive data \citep{rey2005} poses rather stringent
constraints on the aforementioned interpretations for RRATs.  In fact,
the observed spectrum, although the counting statistics is quite low,
appears consistent with a blackbody at $kT\sim 120\pm 40$~eV. Thermal,
soft radiation is not expected from a propeller, in which non-thermal
magnetospheric emission should dominate. PSRs do indeed show a thermal
component in their spectra (e.g. \citealt{pavzav03}) at comparable
temperature. However, this is the case for quite young objects (age
$\approx 10^4$--$10^5\, {\rm yr}$). If RRATs are associated with
pulsars from the death valley, they should be at least one order of
magnitude older and hence much cooler.

Interestingly, the X-ray properties of RRAT J1819-1458 are not in
contradiction with the XDINS picture. The observed temperature is,
within the reported errors, similar to those found in the seven bona
fide XDINSs ($40\la kT\la 100$~eV, see e.g. \citealt{hab04} and
\citealt{zane05}). Would a value of 120~eV be confirmed, this may be
quite naturally explained if this source is younger than the
seven. Actually, the $P/2\dot P$ characteristic age of RRAT J1819-1458
is about 0.11 Myr and this makes it the youngest among the RRATs with
a measured $\dot P$ (for the other two objects ages are 1.8 and 3.3
Myr, see again \citealt{mcl2005}). It should be mentioned in this
respect that in the case of the XDINS RX J0720.4-3125 the
characteristic age turns out to be larger than both the cooling and
the kinematic age \citep{kvk2005}. If this applies to RRAT J1819-1458,
its age may well be below $10^5$ yr.

Given the very limited sample of XDINSs, it is not suprising that no
XDINSs\footnote{Note that the spin period for two of the known XDINSs
is still undetermined.} display a rotational period shorter than $\sim
3$~s: this can be ascribed to the poor probability of discovering
rapidly rotating sources, which spin down very quickly.  We note that
the available sample of RRATs is comparable with that of XDINSs, but
there are RRATs with spin period significantly shorter, down to $\sim
0.4$~s. If RRATs are far away XDINSs, this clearly indicates that the
detectability of the radio bursting activity must be enhanced in
faster spinning neutron stars. Additional support to this
interpretation comes from the fact that two of the RRATs with a
measured period derivative have slightly lower magnetic fields than
the two XDINSs for which a similar measure is available. In fact, a
RRAT with a lower magnetic field spins down more slowly and spends a
longer time in the phase when the detection of radio bursts would be more probable.

\section{Conclusions}
\label{conclus}

The arguments of the previous sections show that, as far as the
birthrate is concerned, XDINSs seem to be a subpopulation of Galactic
neutron stars more abundant than that of PSRs and much larger than
that of SGRs/AXPs. In particular, the birthrate of XDINSs is
consistent with the available estimate made for the birthrate of
RRATs.

Discovery of the radio transient activity from close-by neutron stars
in a survey is a difficult task due to the expected small dispersion
measure towards such objects, which may lead to misinterpret RRAT
candidates as radio interferences.  On the other hand, most of the
known RRATs are too far away for their thermal radiation to be
detectable in the ROSAT All-sky survey above $0.01\, {\rm cts\,
s}^{-1}$.  Radio observations of the known sample of XDINSs (with
sensitivity to their putative bursting activity) are hence necessary
to test the possible link between the two populations.

\section*{Acknowledgments}
We thank Monica Colpi for discussions. The work of SBP was supported
by RFBR grant 04-02-16720 and by the ``Dynasty'' Foundation
(Russia). AP and RT acknowledge financial support from the Italian
Ministry for Education, University and Research through grant
PRIN 2004023189.

%\newpage

\end{document}